\documentclass[aps,floatfix]{revtex4}

\usepackage{amssymb,hyperref,hypcap,
ae,tikz,color}
\usepackage{latexsym,textcomp,slashed,epsfig}
\usepackage[T1]{fontenc}
\usepackage{verbatim}

\usepackage{dcolumn}
\usepackage{bm}
\usepackage{amsmath}
\usepackage{graphicx}
\usepackage{listings}
\usepackage[T1]{fontenc}
\usepackage{xcolor}
\usepackage{lmodern}
\usepackage{listings}

\hypersetup{
	colorlinks=true,
	citecolor=blue,
    linkcolor=magenta,
	}

\begin{document}

\title{{\bf Propagation of Gravitational Waves in Anisotropic Universe}}

\author{\bf{Sucheta Datta and Sarbari Guha}}

\affiliation{\bf Department of Physics, St.Xavier's College (Autonomous), Kolkata 700016, India}

\maketitle

\section*{Abstract}
In this paper, we have considered a toy model of an anisotropic universe and studied the propagation of gravitational waves in such a universe. The consideration of this toy model simplifies the analysis and helps us to illustrate the effects of anisotropy. Incorporating linear perturbations on this anisotropic background, we have considered the synchronous, transverse, traceless gauge conditions and evaluated the perturbations of the Ricci tensor. The energy-momentum tensor is that of a perfect fluid, for which the Einstein's field equations are determined in presence of perturbations. We arrive at the set of linearised Einstein's equations explicitly and find solutions for gravitational waves propagating along the direction of anisotropy. We also study the propagation along a direction perpendicular to the direction of anisotropy. Subsequently we have validated the assumption of the tracelessness of the linear perturbations. Finally we determine the amount of tidal acceleration caused by the propagation of gravitational waves in this background spacetime.

\bigskip

KEYWORDS: Anisotropic universe; Gravitational waves

\section{Introduction}

Gravitational waves (GWs) are `ripples' in the fabric of spacetime generated by massive, accelerating celestial bodies. 
Albert Einstein in 1916 predicted their existence in the context of General Relativity (GR) \cite{EINST1,EINST2}. 
He found that the linearized field equations had solutions representing transverse waves travelling at the speed of light, 
produced by the time-dependence of the mass quadrupole moment of the source. 
The amplitudes of these waves would be small and till 1957, their existence was highly debated. 
The first indirect evidence for the existence of GWs came from the observation of the energy-loss from the binary pulsar system PSR 1913+16, discovered in 1974 by Hulse and Taylor \cite{HT1}. The observed energy-loss \cite{HT2} is in agreement with the theoretical prediction by GR, to within less than 0.3\% error.

Direct detection of GWs involve designing, implementation, and advancement of extremely sophisticated technology as well as mammoth data analysis \cite{GWRVW, GWreRVW}. The first-ever direct detection of GWs was successful on September 15, 2015 from a medium-mass black hole (BH) binary. Observations of GWs from binary BH mergers \cite{GW1,GW2,GW3,GW4,GW4a}, and the single observation of GWs from a neutron star (NS) merger \cite{GW5}, marked the dawn of GW astronomy. The study of GWs will be important for testing the predictions of GR and discriminating between GR and its numerous alternatives.

The linearized theory is a weak-field approximation to GR, where the equations are solved in a nearly flat spacetime. GWs are realised as small perturbations of a time-independent smooth background, whose radius of curvature is much larger than the wavelength of the GWs (`short-wave approximation'). This theory gives an outline of a classical gravitational field whose quantum description would be a massless spin-2 field propagating at the speed of light. On applying the Lorentz gauge together with the transverse-traceless (TT) gauge (outside a sphere enclosing the source), GWs are left with only two degrees of freedom or two independent polarizations: `+' and `$ \times $'. Depending on the non-zero values of these quantities, a circular ring, for instance, may get distorted in any of the two polarization states at a relative orientation of $45^{\circ}$ with each other \cite{WEINBERG, SCHUTZ, HOBSON, LANDAU, PRASANNA, DUNYA}.


Gravitational perturbations in anisotropic universes have been considered by many authors. The propagation of a single GW (specifically, propagation along the symmetry axis of the 2/3, 2/3, -1/3 axially-symmetric Kasner spacetime) in a Bianchi Type-I (B-I) universe was discussed in \cite{HU}. In \cite{MIEDEMA93} GWs have been studied within the context of a general perturbative analysis of the B-I universe and are found to be caused by `non-material' perturbations, which physically implies that they do not locally curve the three-dimensional hyperspaces of constant time.

Unlike the FLRW universe, where the two polarizations of the GW decouple from one another, and each is separately equivalent to a minimally coupled massless scalar field \cite{CHOref}, the two polarizations in an expanding anisotropic universe, as shown in \cite{CHO}, are coupled to one another and the GW acquires an effective mass term because of its spin-2 field and is much more sensitive to the anisotropy than a scalar field. Although the direction of propagation and the polarization of the wave are initially arbitrary, the anisotropic medium itself tends to change both of them.  The anisotropy being time-dependent, the polarization vectors by virtue of the coupling also change with time.

In general, GWs are non-transversal in B-I universe, and even if they are transversal initially, they evolve with time into non-transversal waves. However, the longitudinal components of the GWs can be gauged away in the entire space sections of constant time. This has been demonstrated by Miedema and van Leeuwen \cite{MIEDEMA93} working in the transverse, traceless, synchronous gauge. However, according to Cho and Speliotopoulos \cite{CHO}, the correct choice of gauge will always keep GWs transverse.

The gauge problem plaguing the linear perturbations in cosmological models has been addressed in a number of papers \cite{BARDEEN, STEWART, MUKHANOV, MIEDEMA96}. Based on the gauge-invariant formalism for perturbations on a spherically symmetric space-time developed in \cite{REGGE, SENGUPTA, GUNDLACH, MARTEL}, Clarkson et al. presented  a full system of master equations that represent the general perturbations to LTB space-times in \cite{CLARKSON1}, and discussed the evolution of linear perturbations in an inhomogeneous spherically symmetric background. The complicated equations in that paper was subsequently solved numerically by others \cite{CLARKSON2, BARTELMANN, MEYER}.

Adams et al. \cite{ ADAMS1} have developed a non-perturbative exact formalism to study gravitational radiation of arbitrary polarization, propagating through Bianchi I-VII backgrounds. Choosing a direction of propagation of GWs, say $z$-direction, and inserting z-dependence in the highly symmetric spatially homogeneous Bianchi line element, the symmetry in the chosen direction is broken but preserved in the transverse direction. The wave equation for the $ \psi $ `field' representing the `$+$' polarization is found to be inhomogeneous, with the field propagating through spacetime like a massless scalar field. On  the other hand, the wave equation for the $ \delta $ `field' representing the `$\times$' polarization is homogeneous.
A subsequent paper \cite{ADAMS2} generalizes the property obtained in \cite{ADAMS1} that there exist solutions in which the
inhomogeneity initially dominates the structure of the cosmic singularity and then evolves into gravitational waves propagating
over the more homogeneous background. 

GWs on Minkowski background do not suffer dispersion, scattering or absorption. The `light-like' wave-vector $ k $ satisfies the relation $ w^2=c^2 k^2 $ and both the group velocity and phase velocity are equal to $ c $ \cite{HOBSON}. But GWs, in general, scatter off matter or background inhomogeneities or by non-linear interactions between modes, causing a transfer of energy from lower to higher frequencies \cite{ADAMS1}. The paper \cite{SACCHETTI} shows that the dispersion relation for GWs in FLRW universe is plasma-like in the zero-temperature limit and the phase velocity is smaller than $ c $. 

Many alternative theories predict that GWs propagate at a different speed either by modifying the effective metric in which they propagate, by inducing a mass for the graviton or by introducing higher-order terms in the dispersion relation. Among the various constraints obtained from GW observations, the most stringent ones are the propagation speed equal to the speed of light and the absence of emission of additional polarizations.
Even in FLRW model, the propagation equation in Fourier space is a wave equation with a friction term arising due to the cosmic expansion. 
Here GWs propagate on an effective metric $ \mathcal{G}^{\mu\nu} $ with a  causal structure different from the physical metric $ g^{\mu\nu} $ \cite{EZQref}. The speed can depend on the propagation direction and the frequency, as in the case of massive graviton (mass term $ m_{g} $) \cite{WILL1} or Lorentz-violating modifications (higher-order wave-vector corrections) of the dispersion relation \cite{WILL2}.

Ehlers et al. \cite{PRASANNA1, PRASANNA2} showed that GWs propagating through ideal fluids do not undergo dispersion or dissipation. Generalizing it to the case of non-ideal fluids, Prasanna \cite{PRASANNA3} showed that only the coefficient of shear viscosity causes attenuation of GWs. According to \cite{WEINBERG1}, the issue of treating a particular medium as an imperfect fluid, characterized by coefficients of viscosity and heat conduction, depends on the scales of distance and time of the process under study. 
Moreover, the stochastic background has a directional dependence resulting from the inhomogeneities of the matter distribution \cite{CUSIN}.

In this paper, we introduce a toy model of an anisotropic universe for which the anisotropy is in a particular direction only. Our objective for doing so is to examine the changes that occurs in the propagation of the GWs as compared to the one propagating on a FLRW background. The paper is organised as follows: Sec.II contains the metric for the model and the corresponding Ricci tensor and Einstein's equations. In Sec.III, we incorporate linear perturbations on this background, mention the synchronous, transverse, traceless gauge conditions and evaluate the perturbations of the connection terms and Ricci tensor. Assuming the energy-momentum tensor for a perfect fluid, we write down the Einstein's field equations in presence of perturbations. In Sec.IV we arrive at the set of linearised Einstein's equations explicitly and find solutions for gravitational waves propagating along the direction of anisotropy. We do the same for propagation along a direction perpendicular to the direction of anisotropy in Sec.V. We attempt to justify the assumption of the tracelessness of the linear perturbations in Sec.VI. The tidal acceleration caused by the propagation of gravitational waves is determined in Sec.VII. We conclude with our results in Sec.VIII.

\section{The background metric}
The unperturbed background metric for an anisotropic universe is chosen as
\begin{equation}\label{1}
\overline{g}_{\mu \nu}=
\begin{pmatrix}
1 &0 &0 &0\\0 &-a^{2}(t) &0 &0\\0 &0 &-a^{2}(t) &0\\0 &0 &0 &-a^{2}(t)b^{2}(z)
\end{pmatrix},
\end{equation}
where $a(t)$ is the scale factor with cosmic time $t$, geometrically interpreted as the radius of curvature of the 3-space, and $ b(z) $ represents the anisotropy along the $ z $-direction.

\emph{Throughout this paper} we have assumed geometrized units, i.e., $c=G=1$. The Greek indices run from 0 to 3, and the Latin indices from 1 to 3. In the subsequent analysis, a `dot' is used to denote a derivative with respect to time $ t $, a `prime' for a derivative with respect to spatial direction $ z $, a `tilde' for the derivative with respect to $ x $, and an over-bar represents an unperturbed quantity.

The corresponding non-zero components of the Ricci tensor are:
\begin{equation}\label{2a}
\overline{R}_{00} =  -3 \dfrac{\ddot{a}}{a}, \quad
\overline{R}_{11} = \overline{R}_{22} =  a \ddot{a} +2\dot{a}^2, \quad
\overline{R}_{33} =  (a \ddot{a} +2\dot{a}^2)b^2,
\end{equation}

and the Ricci scalar is
\begin{equation}\label{2e}
\overline{R} 
= -6\dfrac{\ddot a}{a} -6\left( \dfrac{\dot a}{a} \right)^{2}
\end{equation}

\subsection{Energy-momentum tensor and Einstein's field equations}

Let us assume that the matter field is described by a perfect fluid with $u^\alpha$ as the fluid four-velocity, $\rho$ the energy density, $ p $ the pressure, and $\varepsilon = (\rho+p)$ as the heat function per unit volume, so that we have
\begin{equation}\label{8}
T_{\mu\nu}= \varepsilon u_{\mu}u_{\nu}-pg_{\mu\nu} = (\rho+p)u_{\mu}u_{\nu}-pg_{\mu\nu}.
\end{equation}
The trace of the energy-momentum tensor is $ T ={T^{\alpha}}_{\alpha} = \rho-3p $.

The Einstein's field equations for the background metric \eqref{1} are given by
\begin{equation}\label{2f}
\overline{\mathcal{G}}_{\mu\nu} = 8 \pi T_{\mu\nu},
\end{equation}
with the following components: 
\begin{equation}\label{2h}
\overline{\mathcal{G}}_{00} \qquad \textrm{equation :} \qquad \left( \frac{\dot a}{a} \right)^{2} =-\dfrac{8 \pi}{3}\rho\quad ,
\end{equation}
\begin{equation}\label{2i}
\overline{\mathcal{G}}_{11}\quad\text{or}\quad\overline{\mathcal{G}}_{22} \qquad \textrm{equation :} \qquad 2a \ddot{a}+ \dot{a}^{2} =-8 \pi p \quad ,
\end{equation}
\begin{equation}\label{2j}
\overline{\mathcal{G}}_{33} \qquad \textrm{equation :} \qquad (2a \ddot{a}+ \dot{a}^{2})b^2 = -8 \pi p.
\end{equation}

Substituting equation \eqref{2i} in equation \eqref{2j} leads to the constraint:
\begin{equation}\label{2k}
p (b^2 -1) =0
\end{equation}

This indicates that we should have either $ p=0 $, which represents a dust universe, or $ b^2=1 $, which implies the disappearance of anisotropy, thereby reducing our toy model \eqref{1} to the FLRW case. We therefore consider a universe with negligible pressure ($ p\simeq0 $) in our subsequent analysis.

\section{Linear perturbations}

In the linearised theory, the perturbed metric $ g_{\mu \nu} $ includes small perturbations $ h_{\mu \nu} $ over the flat background $ \overline{g}_{\mu \nu} $, where $ h_{\mu \nu} $ is a symmetric (0,2) tensor, considered up to first-order in the calculations. Thus
\begin{equation}\label{3}
g_{\mu \nu}=\overline{g}_{\mu \nu}+h_{\mu \nu},
\end{equation}
with
\begin{equation}\label{5}
h^{\mu \nu} = \overline{g}^{\mu\alpha}\overline{g}^{\nu\beta}h_{\alpha\beta}.
\end{equation}

The synchronous gauge or Lorentz gauge is given by
\begin{equation}\label{6a}
u_{\mu}h^{\mu\nu}=0,
\end{equation}
along with the transverse, traceless condition (TT gauge):
\begin{equation}\label{6b}
\nabla_{\mu} h^{\mu\nu}=0,
\end{equation}
and
\begin{equation}\label{6c}
{h^{\mu}}_{\mu}=0.
\end{equation}
The conditions \eqref{6b} and \eqref{6c} ensure that $h_{\mu \nu}$ does not contain pieces which transform as scalars or vectors \cite{LAZKOZ}.

\subsection{Perturbations of Christoffel connection terms and Ricci tensor}

We follow the steps of \cite{WEINBERG} and \cite{DUNYA} in deriving the first-order perturbations of Christoffel connection terms and Ricci tensor to get
\begin{eqnarray}
 \nonumber
  \Gamma^{\alpha}_{\mu\nu} \simeq  & &\left[ \dfrac{1}{2} \overline{g}^{\alpha\beta} \left(\partial_{\mu} \overline{g}_{\beta\nu} +\partial_{\nu} \overline{g}_{\beta\mu} -\partial_{\beta} \overline{g}_{\mu\nu}\right) \right] +
\end{eqnarray}
\begin{equation}\label{7b}
\hspace{1cm} \left[ \frac{1}{2} \overline{g}^{\alpha\beta} \left(\partial_{\mu} h_{\beta\nu} +\partial_{\nu} h_{\beta\mu} -\partial_{\beta} h_{\mu\nu}\right) - \frac{1}{2} h^{\alpha\beta} \left(\partial_{\mu} \overline{g}_{\beta\nu} +\partial_{\nu} \overline{g}_{\beta\mu} -\partial_{\beta} \overline{g}_{\mu\nu}\right)\right],
\end{equation}
where
\begin{equation}\label{7c}
\overline{\Gamma}^{\alpha}_{\mu\nu} = \left[ \dfrac{1}{2} \overline{g}^{\alpha\beta} \left(\partial_{\mu} \overline{g}_{\beta\nu} +\partial_{\nu} \overline{g}_{\beta\mu} -\partial_{\beta} \overline{g}_{\mu\nu}\right) \right] \end{equation}
are the Christoffel connection terms for the unperturbed metric $ \overline{g}_{\mu\nu} $, and
\begin{equation}
\delta{\Gamma}^{\alpha}_{\mu\nu} = \left[ \frac{1}{2} \overline{g}^{\alpha\beta} \left(\partial_{\mu} h_{\beta\nu} +\partial_{\nu} h_{\beta\mu} -\partial_{\beta} h_{\mu\nu}\right) - \frac{1}{2} h^{\alpha\beta} \left(\partial_{\mu} \overline{g}_{\beta\nu} +\partial_{\nu} \overline{g}_{\beta\mu} -\partial_{\beta} \overline{g}_{\mu\nu}\right)\right]
\end{equation}
denotes the perturbation of the Christoffel connection. Similarly, the Ricci tensor can be expressed as the sum of zeroth (unperturbed) and first-order parts:
\begin{equation}\label{7d}
R_{\mu\nu} = \overline{R}_{\mu\nu} + \delta R_{\mu\nu},
\end{equation}
with the perturbation approximation
\begin{equation}\label{7e}
\delta R_{\mu\nu} \simeq \partial_{\alpha}\left(\Gamma^{\alpha}_{\mu\nu}\right) - \partial_{\mu}\left(\Gamma^{\alpha}_{\alpha\nu}\right).
\end{equation}

\subsection{Energy-momentum tensor and Einstein's field equations in presence of perturbations}

In presence of perturbations, Einstein's field equations
\begin{equation}\label{9a}
R_{\mu\nu}= 8\pi \left[T_{\mu\nu} - \frac{1}{2}g_{\mu\nu}T\right]
\end{equation}
can be split as:
\begin{equation}\label{9b}
\overline{R}_{\mu\nu} + \delta R_{\mu\nu} = 8\pi \left[T_{\mu\nu} - \frac{1}{2}\overline{g}_{\mu\nu}T\right] + 8\pi \left[\delta T_{\mu\nu} - \frac{1}{2}\delta(\overline{g}_{\mu\nu}T)\right].
\end{equation}
Since gravitational waves are non-material perturbations of spacetime \cite{MIEDEMA93}, the first-order perturbations of energy density, pressure and four-velocity of the ideal fluid vanishes for all $ x^i $. Hence $ \delta T_{\mu\nu} =0 $. Moreover,
$$ \delta(\overline{g}_{\mu\nu}T) = h_{\mu\nu}T. $$
Therefore Einstein's equations \eqref{9b} for GWs are given by
\begin{equation}\label{9c}
\overline{R}_{\mu\nu} + \delta R_{\mu\nu} = 8\pi \left[T_{\mu\nu} - \frac{1}{2}(\overline{g}_{\mu\nu} + h_{\mu\nu})T\right].
\end{equation}

\section{Propagation of GW along the $ z $-direction}
When a plane gravitational wave propagates in the spacetime described by the line element \eqref{1} along the $ z $-direction (i.e. along the direction of anisotropy), the perturbation is given by
\begin{equation}\label{10a}
h_{\mu\nu}=
\begin{pmatrix}
0 &0 &0 &0\\ 0 &h_{11} &h_{12} &0\\ 0 &h_{21} &h_{22} &0\\ 0 &0 &0 &0
\end{pmatrix},
\end{equation}
where
\begin{equation}\label{10b}
h_{11}(t,z)= -h_{22}(t,z), \quad
\text{and} \quad
h_{12}(t,z)=h_{21}(t,z).
\end{equation}
From \eqref{5}, we have:
\begin{eqnarray}\label{10d}
h^{11}= \left(a^{-4}\right)h_{11}, \qquad h^{12}=\left(a^{-4}\right)h_{12}, \quad \textrm{and} \quad h^{22}= \left(a^{-4}\right)h_{22}.
\end{eqnarray}

\subsection{Calculation of perturbations of Ricci tensor}

The non-zero perturbations of the Ricci tensor are found to be
\begin{eqnarray}\label{11}
\delta R_{11} =-\dfrac{1}{2}\ddot{h}_{11} + \dfrac{1}{2 a^2 b^2} h_{11}'' - \dfrac{b'}{a^2 b^3}h_{11}' , \\
\delta R_{22} =-\dfrac{1}{2}\ddot{h}_{22} + \dfrac{1}{2 a^2 b^2} h_{22}'' - \dfrac{b'}{a^2 b^3}h_{22}' ,\\
\delta R_{12} = -\dfrac{1}{2}\ddot{h}_{12} + \dfrac{1}{2 a^2 b^2} h_{12}'' - \dfrac{b'}{a^2 b^3}h_{12}' .
\end{eqnarray}

\subsection{Linearised Einstein's equations and their solutions}
The explicit form of the Einstein's equations \eqref{9c} are as follows:
\begin{equation}\label{12a}
R_{00} \quad \textrm{equation :} \qquad
\dfrac{\ddot{a}}{a} =-\dfrac{4\pi}{3}(\rho+3p),
\end{equation}

\begin{equation}\label{12b}
R_{11} \quad \textrm{equation :} \qquad
(a \ddot{a}+2 \dot{a}^2) + \left(-\dfrac{1}{2}\ddot{h}_{11} + \dfrac{1}{2 a^2 b^2} h_{11}'' - \dfrac{b'}{a^2 b^3}h_{11}'\right) =8\pi \left[p+ \frac{a^2}{2}(\rho-3p) -\dfrac{1}{2}(\rho-3p)h_{11} \right],
\end{equation}


\begin{equation}\label{12c}
R_{22} \quad \textrm{equation :} \qquad
(a \ddot{a}+2 \dot{a}^2) + \left(\dfrac{1}{2}\ddot{h}_{11} - \dfrac{1}{2 a^2 b^2} h_{11}'' + \dfrac{b'}{a^2 b^3}h_{11}'\right) =8\pi \left[p+ \frac{a^2}{2}(\rho-3p) +\dfrac{1}{2}(\rho-3p)h_{11} \right],
\end{equation}

\begin{equation}\label{12d}
R_{33} \quad \textrm{equation :} \qquad
(a \ddot{a}+2 \dot{a}^2)b^2 = 8\pi\left[p +\dfrac{a^2 b^2}{2}(\rho-3p)\right],
\end{equation}
and
\begin{equation}\label{12e}
R_{12} \quad \textrm{equation :} \qquad
-\dfrac{1}{2}\ddot{h}_{12} + \dfrac{1}{2 a^2 b^2} h_{12}'' - \dfrac{b'}{a^2 b^3}h_{12}' =0.
\end{equation}
Subtracting equation \eqref{12c} from equation \eqref{12b}, we have
\begin{equation}\label{12f}
\ddot{h}_{11} - \dfrac{1}{a^2 b^2} h_{11}'' + \dfrac{2b'}{a^2 b^3}h_{11}' = 8\pi (\rho-3p)h_{11}.
\end{equation}
Equation \eqref{12e} can be rewritten as
\begin{equation}\label{12g}
\ddot{h}_{12} - \dfrac{1}{a^2 b^2} h_{12}'' + \dfrac{2b'}{a^2 b^3}h_{12}' =0.
\end{equation}
In order to solve equations \eqref{12f} and \eqref{12g}, let us assume the trial solutions:
\begin{eqnarray}
h_{11}(t,z)= Re\left[ \varepsilon_{11} \exp \left\lbrace i(\omega t - k_{11}z)\right\rbrace  \right], \label{13a} \\
h_{12}(t,z)= Re\left[ \varepsilon_{12} \exp \left\lbrace i(\omega t - k_{12}z)\right\rbrace  \right], \label{13b}
\end{eqnarray}
so that a plane gravitational wave of frequency $ \omega $ is characterised by  wave-vectors $ k_{11} $ and $ k_{12} $ in its `+' and `$ \times $'-polarization modes respectively. Substituting  \eqref{13a} in equation \eqref{12f}, and \eqref{13b} in equation \eqref{12g}, and solving for the wave-vectors, the following dispersion relations are obtained:
\begin{equation}\label{14a}
k_{11} 
= -i\dfrac{b'}{b}\pm \dfrac{1}{b} \sqrt{-b'^2 +a^2 b^4 (\omega^2 +8\pi(\rho-3p))},
\end{equation}
\begin{equation}\label{14b}
k_{12} 
= -i\dfrac{b'}{b} \pm \dfrac{1}{b} \sqrt{-b'^2 +a^2 b^4 \omega^2}.
\end{equation}
From these equations it is evident that in this case, the wave vectors depend on the extent of anisotropy, which is determined by $b(z)$. Thus the introduction of anisotropy affects the properties of the GWs, which is expected.

Adding equations \eqref{12c} and \eqref{12b} yields
\begin{equation}
2(a \ddot{a}+2 \dot{a}^2) = 2\left[ 8\pi\lbrace p+ \frac{a^2}{2}(\rho-3p)\rbrace\right],
\end{equation}
and using this in equation \eqref{12d} we arrive at the relation
\begin{equation}\label{12h}
p(b^2 -1) =0.
\end{equation}
as obtained before (equation \eqref{2k}). For $ p\simeq0 $, the wave-vector $ k_{11} $  in \eqref{14a} has contribution from $ a$, $b$ and $\rho $ only.

\section{Propagation of GW along the $ x $-direction}
Considering that a plane gravitational wave propagates in the spacetime \eqref{1} along the $ x $-direction, the perturbation is represented by
\begin{equation} \label{17}
h_{\mu\nu}=
\begin{pmatrix}
0 &0 &0 &0\\ 0 &0 &0 &0\\ 0 &0 &h_{22} &h_{23}\\ 0 &0 &h_{32} &h_{33}
\end{pmatrix}.
\end{equation}
In this case, from \eqref{5} we have
\begin{equation}\label{18a}
h^{22}= \dfrac{1}{a^4}h_{22}, \quad
h^{23}= \dfrac{1}{a^4 b^2}h_{23}, \quad
h^{33}= \dfrac{1}{a^4 b^4}h_{33}.
\end{equation}
The divergenceless condition \eqref{6b} gives rise to two non-trivial relations (on expressing the covariant derivatives in terms of partial derivatives and Christoffel connections of the background metric) :
\begin{equation}\label{19a}
\nabla_{3}h^{32}=0 \hspace{0.4cm}\Rightarrow \partial_{3}h^{23}+ \dfrac{b'}{b}h^{23} =0,
\end{equation}
and
\begin{equation}\label{19b}
\nabla_{3}h^{33}=0 \hspace{0.3cm}\Rightarrow \partial_{3}h^{33}+ 2\dfrac{b'}{b}h^{33} =0,
\end{equation}
whose solutions hint at the nature of $ h_{\mu\nu} $. The solutions are respectively:
\begin{equation}\label{19c}
h^{23} =\dfrac{A}{b},
\end{equation}
\begin{equation}\label{19d}
h^{33} =\dfrac{B}{b^2},
\end{equation}
where $A$ and $B$ are integration constants and can at most be functions of $ z $. Thus the condition \eqref{6b} eliminates the dependence of $ h_{\mu\nu} $ on $ y $ but unlike the previous case retains the $ z $-dependence. 
The traceless condition \eqref{6c} becomes
\begin{equation}\label{20}
{h^{2}}_{2} + {h^{3}}_{3} =0 \hspace{0.4cm} \Rightarrow h_{33}+ b^2 h_{22}=0.
\end{equation}
Therefore, we arrive at the result
\begin{equation}\label{21}
h_{33}(t,x,z) =-b^2 h_{22}(t,x,z) \hspace{1cm} \textrm{and} \hspace{1cm} h_{23}(t,x,z)=h_{32}(t,x,z).
\end{equation}

\subsection{Calculation of perturbations of Ricci tensor}
The non-zero perturbations of the Ricci tensor in this case are as follows:
\begin{equation}\label{22a}
\delta R_{02} = \partial_{3}\left(-\dfrac{1}{2a^2 b^2}\dot{h}_{23} +\dfrac{\dot{a}}{a^3 b^2}h_{23}\right) \; , \qquad
\delta R_{03} = \partial_{3}\left(-\dfrac{1}{2a^2 b^2}\dot{h}_{33} +\dfrac{\dot{a}}{a^3 b^2}h_{33}\right),
\end{equation}
\begin{equation}\label{22b}
\delta R_{12} = \partial_{3}\left(-\dfrac{1}{2a^2 b^2}\tilde{h}_{23}\right) \; , \qquad
\delta R_{13} = \partial_{3}\left(-\dfrac{1}{2a^2 b^2}\tilde{h}_{33}\right),
\end{equation}
\begin{equation}\label{22c}
\delta R_{22} =-\dfrac{1}{2}\ddot{h}_{22}  + \dfrac{1}{2a^2}\tilde{\tilde{h}}_{22} + \dfrac{1}{2 a^2 b^2} h''_{22} - \dfrac{b'}{a^2 b^3}h'_{22} \; , \qquad \qquad
\delta R_{23} = -\dfrac{1}{2}\ddot{h}_{23} + \dfrac{1}{2 a^2} \tilde{\tilde{h}}_{23},
\end{equation}
\begin{equation}\label{22d}
\delta R_{33} =-\dfrac{1}{2}\ddot{h}_{33}  + \dfrac{1}{2a^2}\tilde{\tilde{h}}_{33} - \dfrac{1}{2 a^2 b^2} h''_{33} + \dfrac{2b'}{a^2 b^3}h'_{33} -\dfrac{3b'^2}{a^2 b^4}h_{33} + \dfrac{b''}{a^2 b^3}h_{33}.
\end{equation}

\subsection{Linearised Einstein's equations and their solutions}
The Einstein's equations \eqref{9c} in this case are as follows:
\begin{equation}\label{23a}
R_{00} \quad \textrm{equation :} \qquad \dfrac{\ddot{a}}{a} =-\dfrac{4\pi}{3}(\rho+3p),
\end{equation}
\begin{equation}\label{23b}
R_{11} \quad \textrm{equation :} \qquad  (a \ddot{a}+2 \dot{a}^2) = 8\pi \left[p+ \frac{a^2}{2}(\rho-3p)\right],
\end{equation}
\begin{eqnarray}
\nonumber
R_{22} \quad \textrm{equation :} \, &  &  (a \ddot{a}+2 \dot{a}^2) + \left( -\dfrac{1}{2}\ddot{h}_{22}  + \dfrac{1}{2a^2}\tilde{\tilde{h}}_{22} + \dfrac{1}{2 a^2 b^2} h''_{22} - \dfrac{b'}{a^2 b^3}h'_{22} \right)
\end{eqnarray}
\begin{equation}\label{23c}
 = 8\pi \left[p+ \dfrac{a^2}{2}(\rho-3p) - \dfrac{1}{2}(\rho-3p)h_{22} \right],
\end{equation}
\begin{eqnarray}
\nonumber
  R_{33} \quad \textrm{equation :} \, &  &  (a \ddot{a}+2 \dot{a}^2)b^2 +  \left( -\dfrac{1}{2}\ddot{h}_{33}  + \dfrac{1}{2a^2}\tilde{\tilde{h}}_{33} - \dfrac{1}{2 a^2 b^2} h''_{33} + \dfrac{2b'}{a^2 b^3}h'_{33} -\dfrac{3b'^2}{a^2 b^4}h_{33} + \dfrac{b''}{a^2 b^3}h_{33}\right)
\end{eqnarray}
\begin{equation}\label{23d}
 = 8\pi\left[p +\dfrac{a^2 b^2}{2}(\rho-3p)- \dfrac{1}{2}(\rho-3p) h_{33}\right],
\end{equation}
\begin{equation}\label{23e}
R_{23} \quad \textrm{equation :} \qquad  -\dfrac{1}{2}\ddot{h}_{23} + \dfrac{1}{2 a^2} \tilde{\tilde{h}}_{23}=0,
\end{equation}
\begin{equation}\label{23f}
R_{02} \quad \textrm{equation :} \qquad  \partial_{3} \partial_{0} \left(-\dfrac{1}{2a^2 b^2}{h}_{23}\right) =0,
\end{equation}
\begin{equation}\label{23g}
R_{03} \quad \textrm{equation :} \qquad  \partial_{3} \partial_{0} \left(-\dfrac{1}{2a^2 b^2}{h}_{33}\right) =0,
\end{equation}
\begin{equation}\label{23h}
R_{12} \quad \textrm{equation :} \qquad  \partial_{3} \partial_{1} \left(-\dfrac{1}{2a^2 b^2}{h}_{23}\right) =0,
\end{equation}
\begin{equation}\label{23i}
R_{13} \quad \textrm{equation :} \qquad  \partial_{3} \partial_{1} \left(-\dfrac{1}{2a^2 b^2}{h}_{33}\right) =0.
\end{equation}

Using equation \eqref{23b} in equations \eqref{23c} and \eqref{23d}, we get
\begin{equation}\label{23j}
\ddot{h}_{22}  - \dfrac{1}{a^2}\tilde{\tilde{h}}_{22} - \dfrac{1}{a^2 b^2} h''_{22} + \dfrac{2b'}{a^2 b^3}h'_{22} -8\pi(\rho-3p)h_{22} =0,
\end{equation}
and
\begin{equation}\label{23k}
 \ddot{h}_{33}  - \dfrac{1}{a^2}\tilde{\tilde{h}}_{33} + \dfrac{1}{a^2 b^2} h''_{33} - \dfrac{4b'}{a^2 b^3}h'_{33} + \dfrac{6b'^2}{a^2 b^4}h_{33} - \dfrac{2b''}{a^2 b^3}h_{33} -8\pi(\rho-3p)h_{33} = 16\pi p(b^2 -1).
\end{equation}

Now, using the second equation in \eqref{18a} and eqaution \eqref{19c}, and tacking in the exponential factor, we have
\begin{equation}\label{24a}
h_{23} 
= Re\left[ Aa^{4} b \exp \left\lbrace i(\omega t - k_{23}x) \right\rbrace  \right],
\end{equation}
which when inserted in equations \eqref{23f} and \eqref{23h} reduce them to a single constraint on the integration constant $A$:
\begin{equation}\label{24b}
\partial_{3} \left(\dfrac{A}{b}\right) =0.
\end{equation}

The corresponding solution is
\begin{equation}\label{24c}
A(z)=mb(z),
\end{equation}
where $ m $ is another constant. Therefore
\begin{equation}\label{24d}
 h_{23}(t,x,z)= Re\left[m a^{4} b^{2} \exp \left\lbrace i(\omega t - k_{23}x)\right\rbrace  \right].
\end{equation}
The \emph{real part is implicit in the expressions hereafter}.
Substituting the derivatives of $ h_{23} $ in equation \eqref{23e}, we get
\begin{equation}\label{25a}
 k_{23} =\pm \left[ a^2 \omega^2  - 12\dot{a}^2 - 4a\ddot{a} - 8ia\dot{a}\omega \right] ^{1/2}.
\end{equation}

Similarly, using the third equation in \eqref{18a} and equation \eqref{19d} and tacking in the exponential factor, we have
\begin{equation}\label{26a}
h_{33} 
= Re\left[ Ba^{4} b^{2} \exp \left\lbrace i(\omega t - k_{33}x) \right\rbrace  \right],
\end{equation}
and hence equations \eqref{23f} and \eqref{23h} reduce to a single constraint on the integration constant $B$:
\begin{equation}\label{26b}
\partial_{3}B=0, \quad \textrm{which implies that,} \quad B= \text{constant} = n (\text{say}).
\end{equation}
Therefore
\begin{equation}\label{26d}
 h_{33}(t,x,z)= Re\left[ n a^{4} b^{2} \exp \left\lbrace i(\omega t - k_{33}x) \right\rbrace \right].
\end{equation}

Further, equation \eqref{20} yields
\begin{equation}\label{26e}
h_{22}(t,x)= Re\left[-n a^{4} \exp \left\lbrace i(\omega t - k_{33}x)\right\rbrace  \right].
\end{equation}
So it is evident that $ h_{22} $ has no $ z $-dependence. Let us assume that the wave-vector associated with $ h_{22} $ be $ k_{22} $ and examine whether $ k_{22}$ equals $k_{33} $. 
Substituting the derivatives of $ h_{22} $ and $ h_{33} $ in equations \eqref{23j} and \eqref{23k}, we arrive at two separate equations
\begin{equation}\label{25b}
 k_{22} =\pm \left[ a^2 \omega^2  - 12\dot{a}^2 - 4a\ddot{a} + 8\pi a^2(\rho-3p) - 8ia\dot{a}\omega \right] ^{1/2},
\end{equation}
and
\begin{equation}\label{25c}
n a^2 \left[ b^2(4a\ddot{a} + 12\dot{a}^2 + 8i a\dot{a}\omega -a^2 \omega^2) + b^2 k_{33}^{2}  - 8\pi a^2b^2(\rho- 3p) \right] \exp \left\lbrace i(\omega t - k_{33}x) \right\rbrace =16\pi p(b^2 -1).
\end{equation}
When $ p\simeq0 $, this equation \eqref{25c} results in
\begin{equation}\label{25d}
k_{33} =\pm \left[ a^2 \omega^2 - 12\dot{a}^2 - 4a\ddot{a} + 8\pi a^2\rho - 8ia\dot{a}\omega \right] ^{1/2} ,
\end{equation}
which is exactly same as the equation \eqref{25b} with $ p=0 $. We therefore conclude that in the case of propagation perpendicular to the direction of anisotropy, the wave vectors are independent of the anisotropy factor $b(z)$.

The derivatives of $ h_{33} $ in terms of $ h_{22} $ are 
are now inserted in equation \eqref{23k} to obtain
\begin{equation}\label{28a}
\begin{split}
-b^2 \ddot{h}_{22} + \dfrac{b^2}{a^2}\tilde{\tilde{h}}_{22} - \dfrac{b^2}{a^2 b^2}h''_{22} - \dfrac{4bb'}{a^2 b^2}h'_{22}
- \dfrac{2b'^2}{a^2 b^2} h_{22} - \dfrac{2bb''}{a^2 b^2}h_{22} + \dfrac{4b'b^2}{a^2 b^3}h'_{22} \qquad\qquad\qquad\qquad\qquad\qquad \\
+ \dfrac{4b'(2bb')}{a^2 b^3}h_{22}
- \dfrac{6b'^2 b^2}{a^2 b^4}h_{22} + \dfrac{2b''b^2}{a^2 b^3}h_{22} + 8\pi b^2 (\rho-3p)h_{22} =16\pi p(b^2 -1).
\end{split}
\end{equation}
Multiplying equation \eqref{23j} by $ b^2 $ and adding to equation \eqref{28a} gives
\begin{equation}\label{28b}
h''_{22} + \dfrac{b'}{b^2}h'_{22} = 8\pi a^2 p(b^2 -1).
\end{equation}
Since $ h_{22} $ has no $ z $-dependence, this equation reiterates that $ p\simeq0 $ in presence of the anisotropy factor $ b^2 \neq 1 $ (obtained in equation \eqref{2k}).


We now replace $ mb^2 $ by $ \varepsilon_{23} $, $ -n $ by $ \varepsilon_{22} $, and $ nb^2 $ by $ \varepsilon_{33} $ to write
\begin{equation}\label{29a}
h_{23}(t,x,z)= Re\left[\varepsilon_{23} a^{4} \exp \left\lbrace i(\omega t - k_{23}x)\right\rbrace  \right],
\end{equation}
\begin{equation}\label{29b}
h_{22}(t,x)= Re\left[\varepsilon_{22} a^{4} \exp \left\lbrace i(\omega t - k_{22}x)\right\rbrace  \right],
\end{equation}
and
\begin{equation}\label{29c}
h_{33}(t,x,z)= Re\left[\varepsilon_{33} a^{4} \exp \left\lbrace i(\omega t - k_{33}x)\right\rbrace  \right],
\end{equation}
where $ k_{33}=k_{22} $, and $ \varepsilon_{33}= -b^2\varepsilon_{22} $.

\section{Tracelessness of linear perturbations}
In this section, instead of assuming the condition \eqref{6c} of tracelessness of the linear perturbations $ h_{\mu\nu} $, we try to derive it from the equations in our disposal. If the trace does not vanish, then several $ \delta R_{\mu\nu} $'s no longer vanish and additional terms appear in the Einstein's equations. We will discuss it in both the cases of propagation directions dealt with in this paper.
\subsection{Propagation along $ z $-direction}
Let us suppose that the trace of the linear perturbations is nonzero, and is given by $ h_{11}+h_{22} = h $. Then in particular, we have
\begin{equation}
\delta R_{03}= \partial_{0} \left( \dfrac{1}{2a^2} (h'_{11}+h'_{22}) \right) = \dfrac{1}{2}\partial_{0} \left( \dfrac{1}{a^2} h'\right).
\end{equation}
The corresponding Einstein's equation becomes
\begin{equation}\label{30a}
\dfrac{1}{2}\partial_{0}\partial_{3}\left(\dfrac{1}{a^2}h\right) =0 .
\end{equation}
As $ a(t) $ is a function of time only, the above equation implies that $ h=0 $, which retrieves the traceless condition \eqref{10b}.
\subsection{Propagation along $ x $-direction}
Let $ b^2 h_{22}+h_{33} = h \neq 0 $. One of the several $ \delta R_{\mu\nu} $'s which do not vanish, in particular, is
\begin{equation}
\delta R_{01}= \partial_{0} \left( \dfrac{1}{2a^2}\tilde{h}_{22} + \dfrac{1}{2a^2 b^2}\tilde{h}_{33} \right) = \dfrac{1}{2}\partial_{0} \left( \dfrac{1}{a^2 b^2}\tilde{h} \right).
\end{equation}
The corresponding Einstein's equation is
\begin{equation}\label{30b}
\dfrac{1}{2}\partial_{0}\partial_{1}\left(\dfrac{1}{a^2 b^2}h\right) =0 .
\end{equation}
Once again, as $a=a(t)$ and $b=b(z)$, this result leads us to the traceless condition, i.e. $ h=0 $ as obtained in \eqref{20}.
It is to be noted here that the quantity $ \delta T = \delta(\overline{g}_{\mu\nu}T^{\mu\nu}) =h_{\mu\nu}T^{\mu\nu} $ depends on the trace of $ h_{\mu\nu}$.

\section{Tidal acceleration due to GW}
The effect of a gravitational wave as a tidal force on particles (free or otherwise) can be obtained from the equations of geodesic deviation, which characterises the curvature of spacetime through the induced relative acceleration between the particles.  In locally inertial frame, under non-relativistic approximation, when the four-velocity of the two particles is given by $ u^{\alpha} \equiv (1,0,0,0) $, the relative acceleration of the separation vector $ \chi^{i} $ satisfies the equation
\begin{equation}\label{31a}
\ddot{\chi}^{i}= -R^{i}_{0j0}\chi^{j},
\end{equation}
with the Riemann tensor retaining only the first-order terms in the Christoffel connections:
\begin{equation}\label{31b}
R^{i}_{0j0}= \partial_{j}\Gamma^{i}_{00} - \partial_{0}\Gamma^{i}_{j0}.
\end{equation}

\subsection{Propagation along $ z $-direction}
In this case, $ i $ and $ j $ will be either 1 or 2. The non-zero components of the perturbed Riemann tensor are
\begin{equation*}
R^{1}_{010} 
= -\partial_{0}^2(\log a -\dfrac{1}{2a^2}h_{11}), \quad
R^{1}_{020} 
= -\partial_{0}^2(-\dfrac{1}{2a^2}h_{12}),
\end{equation*}
\begin{equation*}
R^{2}_{010} 
= -\partial_{0}^2(-\dfrac{1}{2a^2}h_{12}) \quad
\text{and} \quad
R^{2}_{020} 
= -\partial_{0}^2(\log a -\dfrac{1}{2a^2}h_{22}).
\end{equation*}
Therefore,
\begin{equation}
\quad 
\ddot{\chi}^{1} = \partial_{0}^2 \left[ (\log a +\dfrac{1}{2}{h^{1}}_1) \right] \chi^{1} + \partial_{0}^2  \left[\dfrac{1}{2}{h^{1}}_2\right] \chi^{2}.
\end{equation}
Assuming that the initial separation $ \chi^{i}(t=0) =\chi^{i}_{0} $ is independent of $ t $ and small compared to $ \partial_{0}^2 ({h^{i}}_{j}) $, the above equation reads
\begin{equation}
\ddot{\chi}^{1} = \partial_{0}^2 \left[ (\log a +\dfrac{1}{2}{h^{1}}_1)\chi^{1} + \dfrac{1}{2}{h^{1}}_2 \chi^{2}  \right],
\end{equation}
for which the solution is
\begin{equation}\label{32a}
 \chi^{1}(t) = (\log a)\chi^{1}_{0} + \dfrac{1}{2}\left( {h^{1}}_1\chi^{1}_{0} + {h^{1}}_2\chi^{2}_{0} \right)   +\chi^{1}_{0}+\chi^{2}_{0}.
\end{equation}

Similarly, for $ \chi^2 $, we obtain the solution
\begin{equation}\label{32b}
 \chi^{2}(t) = (\log a)\chi^{2}_{0} + \dfrac{1}{2}\left( {h^{2}}_2\chi^{2}_{0} + {h^{2}}_1\chi^{1}_{0} \right)   +\chi^{1}_{0}+\chi^{2}_{0}.
\end{equation}

The eigenvalues of $ \chi^{i} $ can be determined from those of the diagonalised perturbation matrix $ \mathsf{h}_{D} $, i.e.
\begin{equation}
\hat{\chi}(t) = \frac{1}{2}\mathsf{h}_{D}(t){\hat{\chi}}_{0}.
\end{equation}
The perturbation matrix is of the form
\begin{equation}
\mathsf{h}(t)=
\begin{pmatrix}
{\varepsilon^{1}}_{1} & {\varepsilon^{1}}_{2} \\ {\varepsilon^{1}}_{2} & -{\varepsilon^{1}}_{1}
\end{pmatrix} e^{\lbrace i(wt-kz)\rbrace},
\end{equation}
having the eigenvalues $ \lambda = \pm \sqrt{({\epsilon^{1}}_{1})^2 + ({\varepsilon^{1}}_{2})^2} $, so that its diagonalised form is
\begin{equation}
\mathsf{h}_{D}(t)=
\begin{pmatrix}
\sqrt{({\varepsilon^{1}}_{1})^2 + {\varepsilon^{1}}_{2})^2} &0\\ 0 & -\sqrt{({\varepsilon^{1}}_{1})^2 + ({\varepsilon^{1}}_{2})^2}
\end{pmatrix} e^{\lbrace i(wt-kz)\rbrace}.
\end{equation}

\begin{itemize}
\item  For `+' polarization, we have $ {\varepsilon^{1}}_{2}=0 $, so that
\begin{equation}
\hat{\chi}^{1}(t) =\dfrac{1}{2} {\varepsilon^{1}}_{1}\exp\lbrace i(wt-kz)\rbrace{\hat{\chi}}_{0}^{1}
\hspace{0.2cm}, \hspace{1cm}
\hat{\chi}^{2}(t) =-\dfrac{1}{2} {\varepsilon^{1}}_{1}\exp\lbrace i(wt-kz)\rbrace{\hat{\chi}}_{0}^{2}.
\end{equation}
\item  For $ `\times $' polarization, $ {\varepsilon^{1}}_{1}=0 $, and therefore
\begin{equation}
\hat{\chi}^{1}(t) =\dfrac{1}{2} {\varepsilon^{1}}_{2}\exp\lbrace i(wt-kz)\rbrace{\hat{\chi}}_{0}^{1}
\hspace{0.3cm},\hspace{1cm}
\hat{\chi}^{2}(t) =-\dfrac{1}{2} {\varepsilon^{1}}_{2}\exp\lbrace i(wt-kz)\rbrace{\hat{\chi}}_{0}^{2}.
\end{equation}
\end{itemize}

\subsection{Propagation along $ x $-direction}
Here, $ i $ and $ j $ will be 2 or 3. The non-zero components of the perturbed Riemann tensor are obtained as
\begin{equation*}
R^{2}_{020} = -\partial_{0}^2(\log a -\dfrac{1}{2a^2}h_{22}), \quad R^{2}_{030} = -\partial_{0}^2(-\dfrac{1}{2a^2}h_{23}),
\end{equation*}
\begin{equation*}
R^{3}_{020} = -\partial_{0}^2(-\dfrac{1}{2a^2 b^2}h_{23})
\quad \text{and} \quad
R^{3}_{030} = -\partial_{0}^2(\log a -\dfrac{1}{2a^2 b^2}h_{33}).
\end{equation*}
Proceeding as before, we arrive at the equation
\begin{equation}
\ddot{\chi}^{2} = \partial_{0}^2 \left[ (\log a +\dfrac{1}{2}{h^{2}}_2)\chi^{2} + \dfrac{1}{2}{h^{2}}_3 \chi^{3}  \right],
\end{equation}
yielding the solution
\begin{equation}\label{33a}
 \chi^{2}(t) = (\log a)\chi^{2}_{0} + \dfrac{1}{2}\left( {h^{2}}_2\chi^{2}_{0} + {h^{2}}_3\chi^{3}_{0} \right)   +\chi^{2}_{0}+\chi^{3}_{0}.
\end{equation}

Similarly, we can find
\begin{equation}\label{33b}
 \chi^{3}(t) = (\log a)\chi^{3}_{0} + \dfrac{1}{2}\left( {h^{3}}_3\chi^{3}_{0} + {h^{3}}_2\chi^{2}_{0} \right)   +\chi^{2}_{0}+\chi^{3}_{0}.
\end{equation}

As before, we can write
\begin{equation}
\hat{\chi}(t) = \frac{1}{2}\mathsf{h}_{D}(t){\hat{\chi}}_{0},
\end{equation}
where the perturbation matrix
\begin{equation}
\mathsf{h}(t)=
\begin{pmatrix}
{\varepsilon^{2}}_{2} & {\varepsilon^{2}}_{3} \\ {\varepsilon^{2}}_{3} & -b^2{\varepsilon^{2}}_{2}
\end{pmatrix} e^{\lbrace i(wt-kx)\rbrace},
\end{equation}
has the eigenvalues
\begin{eqnarray*}
\nonumber \lambda &=& \dfrac{1}{2} \left[ -(b^2 -1){\varepsilon^{2}}_{2} \pm \sqrt{(b^2 -1)^2 ({\varepsilon^{2}}_{2})^2 + 4[({\varepsilon^{2}}_{3})^2 + b^2 ({\varepsilon^{2}}_{2})^2}] \right]  \\
\nonumber     &=& \dfrac{1}{2} \left[ -(b^2 -1){\varepsilon^{2}}_{2} \pm \sqrt{(b^2 +1)^2 ({\varepsilon^{2}}_{2})^2 + 4({\varepsilon^{2}}_{3})^2} \right] ,
\end{eqnarray*}
and its diagonalised form is
\begin{eqnarray}
\nonumber
 \mathsf{h}_{D}(t) &=& a^4 e ^{\lbrace i(wt-kx)\rbrace} \\
 &\times &
\begin{pmatrix}
 \dfrac{1}{2} \left[ -(b^2 -1){\varepsilon^{2}}_{2} + \sqrt{(b^2 +1)^2 ({\varepsilon^{2}}_{2})^2 + 4({\varepsilon^{2}}_{3})^2} \right] &0 \nonumber \\
0 & \dfrac{1}{2} \left[ -(b^2 -1){\varepsilon^{2}}_{2} - \sqrt{(b^2 +1)^2 ({\varepsilon^{2}}_{2})^2 + 4({\varepsilon^{2}}_{3})^2} \right]
\end{pmatrix} . \\
\end{eqnarray}

\begin{itemize}
\item  For `+' polarization, $ {\varepsilon^{2}}_{3}=0 $, and we have
\begin{equation}
\hat{\chi}^{2}(t) =\dfrac{1}{2} a^4{\varepsilon^{2}}_{2}  e^{\lbrace i(wt-kx)\rbrace}{\hat{\chi}}_{0}^{2}
\hspace{0.2cm}, \hspace{1cm}
\hat{\chi}^{3}(t) =-\dfrac{1}{2} a^4 b^2{\varepsilon^{2}}_{2}  e^{\lbrace i(wt-kx)\rbrace}{\hat{\chi}}_{0}^{3}.
\end{equation}
\item  For $ `\times $' polarization, $ {\varepsilon^{2}}_{2}=0 $, which means that
\begin{equation}
\hat{\chi}^{2}(t) =\dfrac{1}{2} a^4{\varepsilon^{2}}_{3} e^{\lbrace i(wt-kx)\rbrace}{\hat{\chi}}_{0}^{2}
\hspace{0.3cm},\hspace{1cm}
\hat{\chi}^{3}(t) =-\dfrac{1}{2} a^4{\varepsilon^{2}}_{3} e^{\lbrace i(wt-kx)\rbrace}{\hat{\chi}}_{0}^{3}.
\end{equation}
\end{itemize}

Hence the non-zero components of tidal acceleration are given by
\begin{equation}\label{34}
\chi^{i}(t) = (\log a)\chi^{i}_{0} + \Sigma_{k} \left[\frac{1}{2}{h^{i}}_k\chi^{k}_{0} + \chi^{k}_{0}\right],
\end{equation}
where $i$ and $k$ take up values $1,2$ in case of propagation along $ z $-direction, and $2, 3$ in case of propagation along $ x $-direction. Summation is carried over $k$.

\section{Conclusions}

In this paper, our analysis is based on a toy model of an anisotropic universe represented by the line element \eqref{1}, which leads us to the conclusion that the proposed metric \eqref{1} represents a universe with negligible pressure (equations \eqref{2k}, \eqref{12h}, \eqref{28b}). For the sake of simplicity, we have assumed that the medium of propagation is in the form of a perfect fluid. From our investigations, we have found that the synchronous, transverse, traceless gauge imposed on the linear perturbations has no issues. Although the assumption of the tracelessness considerably simplifies calculations at several points, yet for this particular metric we find that even if we do away with the condition of traceless gauge, it usually follows from the perturbed Einstein's equations. For both directions of propagation of the GW, the tracelessness of the linear perturbations comes out as a consequence of a particular component of Einstein's equations (equations \eqref{30a} and \eqref{30b} respectively). The traceless condition is slightly modified in the case of propagation perpendicular to the direction of anisotropy. However there is a difference between the two cases for the two directions of propagation of GW considered by us. Unlike the case where the propagation is along $ z $-direction, which is the direction of anisotropy, the divergenceless condition imposes constraints on the nature of the perturbations, in the case where the propagation is along the $ x $-direction.

In both cases, the plane waves are transverse and the polarization states are massless and not coupled to each other. These agree with the results established for FLRW universe \cite{MIEDEMA93, CHO, CHOref}. It is to be noted that equations \eqref{12a} and \eqref{23a} are analogous to temporal component of the Friedmann equations in FLRW spacetime, with a small difference.

The linearised Einstein's equations \eqref{12f} and \eqref{12g} show that the wave equation for `+' polarization is inhomogeneous, while that for `$ \times $' polarization is homogeneous \cite{ADAMS1} when the GW propagates along the $z$-direction. The dispersion relations \eqref{14a} and \eqref{14b} show that the two polarizations have slightly different wave-vectors, the difference being contributed by the energy-momentum tensor. The same holds in the case when the GW moves along the $x$-direction, as is evident from the equations 
\eqref{25a}, \eqref{25b} and \eqref{25d}. 
From the expressions of the wave vectors, it can be inferred that the speed of GWs in our model is lower than $c$, but the amount of lowering depends on both the scale factor $ a(t) $ and anisotropy factor $ b(z) $ when the GWs propagate along the direction of anisotropy, but when the GWs propagate in the direction perpendicular to the direction of anisotropy, the amount of lowering depends only on the scale factor $a(t)$. This is different from what we observe in the case of GWs propagating on a FLRW background, where the speed depends only on $a(t)$. In vacuum, as $ a(t) $ and $ b(z) $ both tend to unit value, one can get back the features of GWs on Minkowski background.

Further, from equations \eqref{29a} and \eqref{29c}, the amplitudes $ \varepsilon_{\mu\nu} $ turn out to be functions of $ z $ in the case of a GW moving along the $x$-direction for the model considered by us. Hence the separation due to tidal acceleration becomes dependent on the anisotropy factor $b(z)$ as well as the scale factor $a(t)$, as we have shown in equations \eqref{33a} and \eqref{33b}. This dependence does not arise when the propagation direction is along $ z $ (equations \eqref{32a} and \eqref{32b}). We therefore conclude that the anisotropy of the medium plays a crucial role on the propagation of gravitational waves.

We intend to generalise the above results in our future works, by considering cosmologies with non-negligible pressure, or with a general anisotropic metric and other features necessary to model realistic situations.

\bigskip


\section{Acknowledgement}
SD acknowledges the financial support from INSPIRE (AORC), DST, Govt. of India (IF180008). SG thanks IUCAA, India for an associateship.

\end{document}